\documentclass[12pt]{iopart}

\usepackage{epsfig,afterpage}
\usepackage{color}
\usepackage{verbatim}
\usepackage{textcomp}
\usepackage{amsfonts,bbold}
\usepackage{amssymb}
\usepackage{graphicx}
\usepackage{esint}
\usepackage{setstack}
\usepackage{amsthm}
\usepackage{dsfont}
\usepackage{epsfig}
\usepackage{subfigure}
\usepackage{iopams}
\usepackage{multirow}
\usepackage{psfrag}

\def\beq{\begin{equation}}
\def\eeq{\end{equation}}
\def\bea{\begin{eqnarray}}
\def\eea{\end{eqnarray}}
\def\bq{\begin{quote}}
\def\eq{\end{quote}}
\def\ben{\begin{enumerate}}
\def\een{\end{enumerate}}
\def\bit{\begin{itemize}}
\def\eit{\end{itemize}}

\def\eqref{\eref}

\newcommand{\tfrac}{\textstyle\frac}

\def\ssigma{\tilde{\sigma}}

\begin{document}

\title{Wigner function for a particle in an infinite lattice}

\author{M Hinarejos$^1$, A P\'erez$^1$ and M C Ba\~nuls$^2$}
\address{$^1$Departament de F\'{\i}sica Te\`orica and IFIC,
Universitat de Val\`encia-CSIC,
 Dr. Moliner 50, 46100-Burjassot, Spain}
\address{Max-Planck-Institut f\"ur Quantenoptik, Hans-Kopfermann-Str.
1, Garching, D-85748, Germany.}

\begin{abstract}
We study the Wigner function
for a quantum system with a discrete, 
infinite dimensional Hilbert space, such as a spinless particle moving on a 
one dimensional infinite lattice. 
We discuss the peculiarities of this scenario and of the 
associated phase space construction, propose a meaningful definition of the Wigner function in this case, and
characterize the set of pure states for which it is non-negative.
We propose a measure of non-classicality for states in this system 
which is consistent with the continuum limit.
The prescriptions introduced here are illustrated by applying them to localized and Gaussian states, and
to their superpositions. 
\end{abstract}

\maketitle

\section{\label{sec:introduction}Introduction}

The formalism of Wigner functions and the 
formulation of quantum mechanics in phase space
have been used since the early days of quantum physics. 
Originally motivated by the attempt to describe quantum effects in thermal ensembles, 
various quasi-probability distribution functions have been developed
and applied to many different fields in quantum physics, 
as alternative formalisms which provide useful computational tools and
facilitate physical insight into the quantum nature of states~\cite{Hillery1984121,ferrie11review}.

The first quasi-probability distribution function was introduced by Wigner in 1932~\cite{PhysRev.40.749}, in the context of statistical mechanics, to study quantum corrections to the thermodynamic equilibrium properties.
In analogy to the classical situation, in which a state can be completely described in terms of its phase-space density,
a quantum state can also be entirely characterized by its Wigner function,
and the expectation values of all observables can be computed as a sum over the whole phase-space weighted by this function.
In contrast to the classical case, in the quantum scenario 
probability distributions cannot be defined simultaneously over position and momentum. Thus the Wigner function is not a true probability distribution, 
as becomes apparent in the fact that it can adopt negative values. Instead, it can be interpreted as a quasi-probability distribution whose marginals reproduce the true probability distributions over single observables.
Operators and dynamics can also be accommodated in the phase-space picture~\cite{moyal49},
so that quantum mechanics can be entirely formulated in this framework.


The Wigner function and other phase-space representations have found application to many different physical problems~\cite{Hillery1984121}, 
including quantum optics, statistical mechanics, hydrodynamics, nuclear theory and quantum field theory.
In particular, in the field of quantum optics, the phase-space descriptions of quantum states have found extensive application. 
Specially interesting from the experimental point of view is the ability to reconstruct the Wigner function (and thus the quantum state)
from measurements of the electromagnetic field quadratures, thus making it a very powerful tool for state tomography~\cite{bertrand87,vogel89,citeulike:4313181}.
Remarkably, the fact that the Wigner function is not positive definite has itself a practical use, since 
the volume of its negative part can be used as a measure of non-classicality of the state~\cite{kenfack}.




In the last two decades, the interest in quantum information processing systems boosted the 
 generalization of Wigner functions to quantum systems in finite dimensional Hilbert spaces.
Although an early approach was pioneered by Stratonovich~\cite{stratonovich},
who introduced a spherical, continuous phase space for a spin particle, 
more recent definitions target discrete phase spaces.
 The first such generalization was proposed by Wootters in 1987~\cite{ Wootters1987} for prime dimensional systems,
and later generalized to any power of primes in~\cite{PhysRevA.70.062101}. A different construction was 
followed in~\cite{leonhardt95prl,PhysRevA.53.2998,miquel02qc} which could cope with any dimension of the 
Hilbert space at the expense of enlarging the size of the phase space grid. The concept of negativity has also been studied in 
this context~\cite{galvao05speedup,cormick06class,gross07prime} and connected to beyond-classical features of quantum algorithms.

In this work we focus on a scenario not covered by the previous literature, namely that 
of a quantum system with an infinite dimensional but discrete Hilbert space.
The most immediate example would be a spinless particle moving on a one-dimensional lattice.
We present a definition of the phase space and the Wigner function for this situation, which connects to both the features of the 
discrete constructions mentioned above and the proper continuum limit. 


The rest of the paper is organized as follows.
In Sect.~\ref{sec:review} we review the construction of the continuous
and discrete Wigner functions used in the literature, and comment on
the different features of each case. Sect.~\ref{sec:definition}
presents our definition of the Wigner function for the infinite
discrete quantum system, sketches the proof of its main properties,
and shows that it reproduces the correct continuous limit.
In Sect.~\ref{sec:negativity} we introduce a measure of
non-classicality of the states which can be computed from the Wigner
function.
Our definitions are illustrated in Sect. ~\ref{sec:examples} 
with an explicit calculation of the Wigner function and the 
non-classicality for several examples.
Finally, in
Sect.~\ref{sec:discussion} we discuss the utility of this definition,
and how it can be applied to more general settings, for instance, to
the case of a particle with spin moving on the lattice, or to several
particles.

\section{\label{sec:review}Continuous and discrete Wigner functions}

The phase spaces of continuous and discrete quantum systems turn out to have striking differences. Defining a Wigner function for a discrete case, thus, requires more than a simple discretization of the continuum equations.
There have been several prescriptions proposed for this kind of systems. 
In this section, after reviewing the main characteristics of the continuous Wigner function that a discrete version should respect, 
we summarize the different approaches that have been proposed,
and their connection to the continuous definition.

\subsection{Continuous case}

For a quantum one-dimensional system, the Wigner function can be written~\cite{PhysRev.40.749}
\begin{equation}
W_{\mathrm{c}}(x,p)=\frac{1}{\pi} \int_{-\infty}^{\infty} dy  \langle x+y |\rho| x-y \rangle \ e^{-i 2 p y},
\label{eq:wfcont}
\end{equation}
where $\rho$ is the density matrix of the system, and $|x\rangle$ represents the eigenbasis of the 
position operator, $\hat{X}$~\footnote{We use natural units, such that $\hbar=1$.}.

It is also possible to define the Wigner function axiomatically~\cite{bertrand87}.
The fundamental properties that it must satisfy are usually formulated as follows. 
\begin{enumerate}
\item{\label{prop1}Reality:}
the Wigner function is real.
\item{\label{prop2}Projection:}
the integral of the Wigner function along any direction, $(\alpha,\beta)$, in phase space, yields the probability distribution for the outcomes of measuring the observable $\alpha \hat{X}+\beta \hat{P}$, being $\hat{P}$ the momentum operator. In particular, then, the marginal distributions for position and momentum can be obtained, respectively, by $\int dp W(x,p)=P(x)$ and $\int dx W(x,p)=P(p)$.~\footnote{In the rest of the paper, unspecified integral (or sum) limits will be understood as extending over the whole range of the integrated (summed) variable.}
\item{\label{prop3}Inner product:}
the inner product of two states, given by their density operators, $\rho_1$ and $\rho_2$, can be computed from their Wigner functions as
\beq
\mathrm{tr} \left ( \rho_1 \rho_2 \right )= 2 \pi \int dx dp W_1(x,p) W_2(x,p).
\eeq
The expectation value of any operator, $\hat{O}$, can be also computed from its Wigner representation, $W_O(x,p)$.
\end{enumerate}


The Wigner function can also be constructed from the \emph{phase-point operators}, 
defined for all points in the phase space as
\begin{equation}
A(x,p)\equiv \frac{1}{\pi}{\cal D}(x,p)\Pi{\cal D}(x,p)^{\dagger},
\label{eq:Acont}
\end{equation} 
where ${\cal D}(x,p)$ are displacement operators and $\Pi$ is the parity reflection. 
The phase-point operators form a complete set,  spanning all Hermitian operators. In particular, the Wigner function corresponds to 
coefficients of the density matrix in this basis,
\begin{equation}
W(x,p)=\mathrm{tr} \left( \rho A(x,p)\right ),
\end{equation}
so that the full state can be reconstructed by 
\begin{equation}
\rho= \int dx dp W(x,p) A(x,p). 
\end{equation}

Properties equivalent to~(\ref{prop1})-(\ref{prop3}) can be formulated for phase-point operators, leading to the same definition of the phase space.
According to these properties the operators $A(x,p)$ should be Hermitian and satisfy an orthogonality condition, and integrating $A(x,p)$ along a line in phase space must yield a projector.

\subsection{Discrete finite case}

\begin{figure}[floatfix]
\psfrag{k}[c][l]{$k$}
\psfrag{l}[c][c]{} 
\psfrag{m}[c][c]{$m$}
\psfrag{n}[c][c]{} 
  \includegraphics[width=.65\columnwidth]{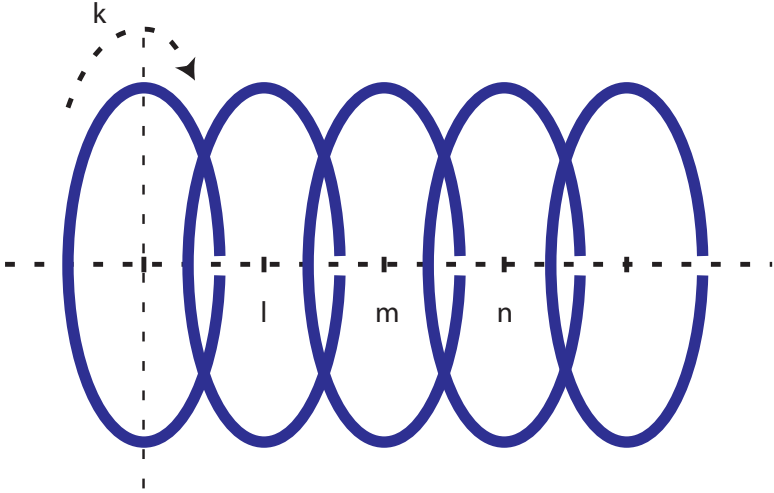}
\caption{Graphical representation of the phase space for an infinite one-dimensional lattice. The \emph{momentum-like} coordinate is continuous and periodic, $k\in[-\pi,\pi[$, and the \emph{position-like} coordinate is discrete, labelled by integer values, $m$.}
 \label{fig:discretePS}
\end{figure}

A valid generalization of the Wigner function to the case of a discrete Hilbert space involves generalizing the concept of phase space and the definition of phase space operators.
Several approaches have been proposed in the literature for the case of a finite dimensional, periodic Hilbert space. 
Here we briefly describe the two main alternatives, emphasizing their relation to the continuous case, and we establish 
the basis for our definitions (see~\cite{vourdas04review,ferrie11review} for more comprehensive reviews).

Wootters~\cite{Wootters1987} generalized the definition of the Wigner function to discrete periodic Hilbert spaces of prime dimension, $N$.
In Wootter's original construction the phase space was a two dimensional $N\times N$ array, indexed by integers. Complete sets of parallel lines in this phase space, or  \emph{striations}, which are defined using arithmetics modulo $N$, were associated to projective measurements.
For more general cases, such as power of primes dimensions or composed systems with both discrete and continuous degrees of freedom, the phase space could be constructed as a Cartesian product of the fundamental pieces.
A related, more general construction, valid for systems whose dimension is an integer power of a prime number, was put forward in~\cite{PhysRevA.70.062101}. In the general construction the discrete phase space has also size $N\times N$, and was labelled by a finite field. To give a physical interpretation to the discrete phase space, each line is assigned to a pure quantum state. The set of all lines parallel to a given one corresponds to an orthogonal basis, and the distinct sets are mutually unbiased basis~\cite{wootters86mub}. This assignment of states to lines determines the particular definition of the Wigner function for the system, which is therefore not unique. 
Although closely connected to quantum information concepts and useful to describe systems of $n$ qubits~\cite{PhysRevA.72.012309,cormick06class}, 
the lack of a unique physical interpretation and the restriction to dimensions which are powers of primes
makes this approach less appropriate for the kind of system we want to describe.

Leonhardt~\cite{PhysRevA.53.2998} introduced another definition, more closely connected to the continuous construction, in which the labels of the phase space axis could be connected to discrete position and momentum basis of the physical system. For the case of an odd dimensional system, the discretized version of the continuous definitions is enough to obtain a valid definition of the Wigner function.
However, in the case of even dimensional Hilbert spaces, the naive discretization does not suffice to guarantee a Wigner function with the desired properties. Instead, half-odd labels had to be introduced between the integer points of the phase space axis, so that the size of the grid has to be increased  to $2N\times 2N$ (see also~\cite{1751-8121-44-34-345305} for a discussion). 
A similar approach was pursued in~\cite{miquel02qc}, where the construction followed from the definition of discrete phase-point operators and was then applied to the analysis of quantum algorithms~\cite{PhysRevA.68.052305}.
In~\cite{paz02mixed}, this approach was also combined with Wootters' prescription to compose degrees of freedom and employed for the study of quantum teleportation.
Our approach builds up on this construction, 
easily connected to the physical interpretation of the continuous phase space.

\section{\label{sec:definition}Definition in the infinite discrete lattice}

We consider here a single particle moving on a discrete one-dimensional lattice, with inter-site spacing $a$.
We can define a discrete position basis, given by orthonormal states  $|n\rangle$, with $n\in \mathbb{Z}$.
Its Fourier transform defines then a quasi-momentum basis, 
$|q\rangle=\sqrt{\frac{a}{2 \pi}}\sum_n e^{i q n a}|n\rangle$, 
which can be restricted to the first Brillouin zone, $q\in[-\frac{\pi}{a},\frac{\pi}{a}[$.

Unlike the discrete cases considered above, the Hilbert space of this system is not periodic and has infinite dimension.
The continuum limit is recovered as $a\rightarrow0$, and $\frac{1}{\sqrt{a}}| n \rangle\rightarrow |x=n a\rangle$. 
We may require that the Wigner function for this system,
besides fulfilling the defining properties, also reproduces in that limit the usual one for a particle in one dimension,
\eqref{eq:wfcont}.

In~\cite{PhysRevA.53.2998,miquel02qc} the problems in the direct discretization of Eq. \eqref{eq:Acont} were connected to the fact that, in the case of even dimensions, such a definition does not generate enough independent operators.
We may thus wonder whether the fact of having an infinite dimensional system is enough to solve this problem.
However, the direct discretization of \eqref{eq:Acont}, in spite of producing an infinite number of operators, does not suffice either in this case to obtain a Wigner function that fulfills the desired properties in the case of interest.

Here we follow closely the construction of~\cite{miquel02qc}, starting with a definition of the discrete phase space and the associated phase-point operators that then produce the Wigner function.
 
A direct discretization of the phase-point operators leads to
\begin{equation} 
A_{\mathrm{direct}}(m,k)=\frac{1}{2 \pi}U^{2 m}\hat{\Pi}V(-\frac{2 k}{a}) e^{i 2 k m},
\end{equation} 
where $U$ is the discrete translation operator, shifting the lattice by one site, $U^m|n\rangle=|n+m\rangle$, and $V(q)$ is the continuous momentum translation, defined by its action on the momentum basis as $V(q')|q\rangle=|q+q'\rangle$.
Notice the hybrid character of the phase space in this case, with discrete and unbounded values of $m$ and continuous, periodic $k\in[-\pi,\pi[$. 
It is easy to see that $A_{\mathrm{direct}}$ has periodicity $\pi$ in the momentum coordinate, $A_{\mathrm{direct}}(n,k\pm \pi)=A_{\mathrm{direct}}(n,k)$.
The Wigner function following from these phase-point operators does not fulfill the defining properties.
In particular, summing over positions does not produce the correct marginal. Instead,
\begin{equation}
\sum_n W_{\mathrm{direct}}(n,k) \propto \left\langle \tfrac{k}{a} | \rho | \tfrac{k}{a} \right\rangle +\left\langle \tfrac{k+\pi}{a} | \rho | \tfrac{k+\pi}{a} \right\rangle.
\end{equation}
As a consequence, the resolution in $k$ is not enough to retrieve all the information on the state from the Wigner function (see~\cite{PhysRevA.53.2998,1751-8121-44-34-345305} for a discussion of this effect).

The problem does not appear if we integrate over the momentum coordinate, and is thus an effect arising purely from the discrete character of the position basis.
It is not surprising, then, that the strategy in~\cite{miquel02qc}, consisting in doubling the number of points in the phase space, serves us to define also here an appropriate set of phase-point operators. In our case, the doubling should only affect the position coordinate, and is equivalent to adopting the definition
\begin{equation}
A(m,k)=\frac{1}{2 \pi}U^m \hat{\Pi}V(-\tfrac{2 k}{a}) e^{i m k}.
\label{eq:Aop}
\end{equation} 
In the position basis, the phase-point operators can then be written as $A(m,k)=\frac{1}{2 \pi}\sum_n |m-n\rangle \langle n|e^{-i (2 n-m) k}$, and 
the Wigner function for our system reads
\begin{equation}
W(m,k)\equiv \mathrm{tr}\left [ \rho A(m,k)\right ]=\frac{1}{2\pi}\sum_n \langle n|\rho|m-n\rangle e^{-i (2 n -m) k}.
\label{eq:WFdef}
\end{equation}
This corresponds to a phase space with the structure depicted in \fref{fig:discretePS}, where the $m$ coordinate takes integer values, whereas $k$ is continuous and periodic, taking values in $[-\pi,\pi[$.
Notice that ($m$, $k$) cannot be directly interpreted as discrete positions, and quasimomenta. Instead, they have to be understood as labels of the phase space points. To make the distinction clear, we reserve symbols $m$ and $k$ for the phase space and $n$, $q$ for the position and quasimomentum states of the lattice.

Although the periodicity of $A(m,k)$ is now $2 \pi$,
it is important to notice that not all the operators are independent. Indeed, $A(m,k \pm \pi)=(-1)^m A(m,k),$ from which it follows
\begin{equation}
W(m,k \pm \pi)=(-1)^m W(m,k).
\label{eq:phases}
\end{equation}

It is easy to check that the definition \eqref{eq:WFdef} fulfills the main properties we require from a valid Wigner function.
In particular it is real, as follows from the Hermiticity of \eqref{eq:Aop}. 
The inner product property~(\ref{prop3}) is also easy to check, given operators $\hat{A}$ and $\hat{B}$,
\beq
2 \pi \sum_{m=-\infty}^{\infty}  \int_{-\pi}^{+\pi} dk  W_A(m,k) W_B(m,k)
= \mathrm{tr} \left ( \hat{A}\hat{B}\right ),
\label{eq:innerprod}
\eeq

In a very similar way, we obtain the explicit expression of the density operator in terms of the Wigner function,
\begin{equation}
\rho=2\pi\sum_m\int_{-\pi}^{+\pi} dk W(m,k) A(m,k).
\label{eq:rhoWF}
\end{equation}

Due to the relation \eqref{eq:phases},
the orthogonality relation between phase-point operators adopts the following form,
\bea
\mathrm{tr}\left [ A(m_1,k_1)  A(m_2,k_2) \right ] &=&\frac{1}{4 \pi} \delta_{m_1 m_2} \left [ \delta(k_1-k_2) 
 +(-1)^{m_1}\Theta(k_2)\delta(k_1-k_2+\pi)
 \right.\nonumber
 \\
&&\left.
+(-1)^{m_1}\Theta(-k_2)\delta(k_1-k_2-\pi)\right],
\label{eq:Aorth}
\eea
where $\Theta(k)$ is the Heavide step function.
To obtain this relation we made use of $\sum_n e^{i n k}=2 \pi \sum_r \delta(k+2\pi r)$, where the sum runs over all  $r\in\mathbb{Z}$.
Eq. \eqref{eq:Aorth} reflects the fact that operators associated to phase space points whose $k$ coordinate is shifted by $\pi$
are not independent, but differ only in a phase.

We may also compute the marginal distributions of \eqref{eq:WFdef}, and obtain
\begin{equation}
\sum_{m=-\infty}^{+\infty} W(m,k) = \tfrac{1}{a} \langle \tfrac{k}{a}|\rho| \tfrac{k}{a} \rangle,
\end{equation}
and
\begin{equation}
\int_{-\pi}^{+\pi} dk W(m,k)=
\sum_n \delta_{m,2n}\langle n |\rho| n \rangle.
\end{equation}
The last equations make evident the distinction between the coordinates of the momentum space points, $m\in \mathbb{Z}$, $k\in[-\pi,\pi[$, and 
the position and quasimomentum bases, $n,$ $q$.
The $k$ coordinate is adimensional and does not directly represent a momentum value, but is connected to $q=k/a$.
The \emph{spatial} label $m$ in phase-space is only connected to a discrete position, $s$, for 
even values , $m=2 s$, while the odd values of $m$  
are analogous to the odd half-integer phase space grid points in~\cite{PhysRevA.53.2998,miquel02qc}.  

Keeping these considerations in mind, we can take the continuum limit that transforms our discrete lattice into real space. This limit is attained by letting $a\rightarrow 0$, with $n a \rightarrow x\in \mathbb{R}$.
With this prescription, we can easily see that the continuum limit of Eq. \eqref{eq:WFdef} yields (up to a proportionality factor) the proper
continuum Wigner function,
\begin{eqnarray}
W(m,k) &\underset{a\to 0}{\longrightarrow} &
\frac{1}{2}W_{\mathrm{c}}(y=\frac{ma}{2},q=\frac{k}{a}) \nonumber \\
&&=\frac{1}{2 \pi}\int_{-\infty}^{+\infty}d z \langle \frac{ma}{2}+z |\rho|\frac{ma}{2}-z \rangle e^{-i 2z\frac{k}{a} }
,
\label{eq:wfcontlim}
\end{eqnarray}
as can be checked from the definition~\eqref{eq:wfcont} after a simple change of variable.
Together with the discussion above, this result shows how the proper continuum limit is attained in the phase space coordinates. Indeed, as the spacing is decreased, $m \frac{a}{2} \to x$ and $\frac{k}{a}\to q$. The Wigner function is a quasi-probability distribution, and the
physically meaningful quantities are given by integrals over the phase space.
The measure of the integration must be modified according to this change of
variables, so that we obtain the correspondence, for the integral over any region of the phase space,
\beq
\sum_m \int dk W(m,k)  \underset{a\to 0}{\longrightarrow} \int dy \int dq W(y,q).
\label{eq:contdiff}
\eeq

\section{\label{sec:negativity}Non-classicality of states: negativity of the Wigner function}

The fact that the Wigner function is not positive definite over the phase space is interpreted as a quantum feature, since it follows from the incompatibility of 
quantum observables. 
This property has been  applied to separate quantum states from classical ones.
In the continuous case, it is known that the only pure states with non-negative Wigner function are Gaussian states~\cite{hudson74}. The classification is not so clear for mixed states, where nevertheless some bounds are known for states with positive Wigner function~\cite{mandilara09}.
From a quantitative point of view, the volume of the negative part of the Wigner function can be used as a measure of non-classicality~\cite{kenfack}. More recently, it has been shown that the smallest distance to a state with positive Wigner function can
also be used to measure the non-classicality of a state, without needing full tomography~\cite{PhysRevLett.106.010403}. 

In the context of discrete systems the negativity of the Wigner function has also been
explored, but the different prescriptions discussed in the previous section lead to 
different conclusions. For the direct discretization, a discrete
version of Hudson's theorem was proven by
Gross~\cite{gross06hudson,gross07prime} for the case of a Hilbert space with odd dimension. In that case
the only pure states with non-negative Wigner functions are stabilizer
states. 
For the class of discrete Wigner functions defined as in~\cite{PhysRevA.70.062101}
a characterization was given in~\cite{galvao05speedup},
where the set of non-negative states was identified with the 
convex hull of stabilizer states when the Hilbert space dimension was small.
Another scenario studied in the literature, which closely relates to our construction, is that of a pair of quantum variables, 
angular momentum and angle, and their associated phase space~\cite{rigas10,rigas11}
which has the same mixed discrete-continuous structure of \fref{fig:discretePS}. For that case, it has been shown that the only
states with non-negative Wigner functions are those of well defined angular momentum. 

For the situation studied in this paper, a similar reasoning to that in~\cite{rigas10}
leads to the conclusion that
 a pure state has a non negative Wigner function if and only if
it is a state of well defined position in the lattice,
i.e. the components of the state vector in the position basis are given by a delta function (see the Appendix for details).

With our definition, however,  phases \eqref{eq:phases} imply that
any state with a non-vanishing Wigner
function on some phase space point with odd-valued position-like
coordinate, $W(m=2s+1,k)\neq 0$, will necessarily have a contribution
of opposite sign at points $(m,k\pm \pi)$. 
These signs are fundamental in order
to ensure that the Wigner function reproduces the momentum and
position probability distributions, but are not related to the
\emph{quantumness} of the different states.
Therefore a
naive calculation of the volume of the negative part of the function,
i.e. applying the discrete version of the definition
in~\cite{kenfack}, will not be a valid measure of
non-classicality, as it would result in a non-vanishing value even for
states expected to be \emph{classical}, such as the discrete version
of Gaussian states.

A similar phenomenon has been observed in different contexts.
In the field of signal analysis, where Wigner functions have also been 
widely employed, the discrete time Wigner distribution shows similar features, which are related to aliasing~\cite{peyrin86}, and various alternative definitions have been proposed to construct alias-free distributions, and to allow a reconstruction of the continuum time signal from a discrete sample.
In the context of finite dimensional quantum systems, a proposal for a \emph{ghost free} Wigner function was put forward in~\cite{arguelles05}.
In all such cases, the negative values of the Wigner function 
respond to the very structure of the discretized
phase space and not to the features of the state or the signal. 
We would thus like to define a new quantity which 
serves to estimate non-classicality of states in our system, and allows for a 
connection to the well-defined continuum limit.
In particular, we expect that this non-classicality measure vanishes for all Gaussian states,
so as to reproduce the well-known continuum limit, and that it 
does not include the spurious negative parts from the extended phase
space.

The definition \eqref{eq:WFdef} leads to a discrete Wigner function which contains two \emph{images},
one in each half of the momentum domain.
According to \eqref{eq:phases}, they have the same magnitude, but on odd position-like 
coordinates $m$, their sign is reverted,
as can be seen in \fref{fig:gauss} for the case of a pure Gaussian state.
Although in the continuum limit  \eqref{eq:WFdef} reduces to the original expression for the Wigner function,
and the second, \emph{ghost} image disappears, we would like to
have a quantity that characterizes the \emph{non-classicality} of discretized states. In particular, we
require that the criterion is consistent with that for the
analogous continuous states, in the cases when such exist.

The so-called ghost image exhibits alternating signs between even and odd space-like coordinates 
(see, for instance \fref{fig:gaussDispl}), while the regular image is smooth. 
We would thus like to use as a measure of the non-classicality the negativity restricted to the
regular image. However its position in the phase space plane is not fixed, but changes with 
momentum shifts. 
A momentum displacement, $q_0$, translates into a displacement $q_0 a$ in the $k$ coordinate. 
As the lattice spacing vanishes the regular image lies on the central region of the phase space, while 
the ghost image is pushed towards the edge, which in the continuum is mapped to infinity.

Instead of trying to locate the regular image, so as to restrict the sum to the corresponding phase-space region, we may apply a filter that 
eliminates the spurious sign oscillations from odd values of $m$, and effectively produces two copies of the regular image.
We thus define the following quantity,
\begin{equation}
\eta(\rho)\equiv  \sum_{m=-\infty}^{+\infty}\int_{-\pi}^{+\pi} dk \left[|W^{(s)}(m,k)|-W^{(s)}(m,k)\right],
\label{eq:neg}
\end{equation}
where $W^{(s)}$ is the result of filtering out the sign oscillations for odd $m$.
If the filtering is perfect, in the continuum limit $\eta(\rho)$ will yield twice the negativity of the Wigner function as defined in~\cite{kenfack}.

Different filtering operations can be tried to this aim. In particular, we propose to use a  sign-averaged Wigner function, defined by
\begin{equation}
W^{(s)}(m,k)\equiv \left\{
\begin{array}{l l}
W(m,k) & m\    \mathrm{even}\\
\chi(m,k) |W(m,k)| & m\   \mathrm{odd},
\end{array}
\right.
\label{eq:defWtilde}
\end{equation}
with $\chi(m,k)=\mathrm{sign}[2\ \mathrm{sign}(W(m-1,k))+\mathrm{sign}(W(m,k))+2\ \mathrm{sign}(W(m+1,k))]$, 
i.e. the even components are unchanged, and
the sign of the odd ones is corrected according to a majority criterion that takes into account 
the sign of the two closest neighboring even points
\footnote{The factor $2$ takes care of the situation when one of the adjacent even points has vanishing $W(m,k)$.}.
This produces approximately two copies of the regular image (see \fref{fig:gaussWtilde}), so that $\eta(\rho)$ is equivalent to twice the negativity
restricted to the half space where this image is supported.

\section{\label{sec:examples}Particular cases}

To illustrate the definitions introduced in the previous sections, we
explicitly compute here the Wigner functions and negativities for
several pure states.

\subsection{Localized state}

We may consider the simplest case, in which the position of the
particle is well defined,
$|\Psi_{\delta}\rangle=|n_0\rangle$, so that, in position basis,
$\langle n |\Psi_{\delta}\rangle=\delta_{n\, n_0}$.
The Wigner function can then be computed exactly,
\begin{equation}
W_{\delta}(m,k)=\frac{1}{2\pi}\delta_{m\, 2n_0}.
\label{eq:WFdelta}
\end{equation}

\begin{figure}
\includegraphics[width=.6\columnwidth]{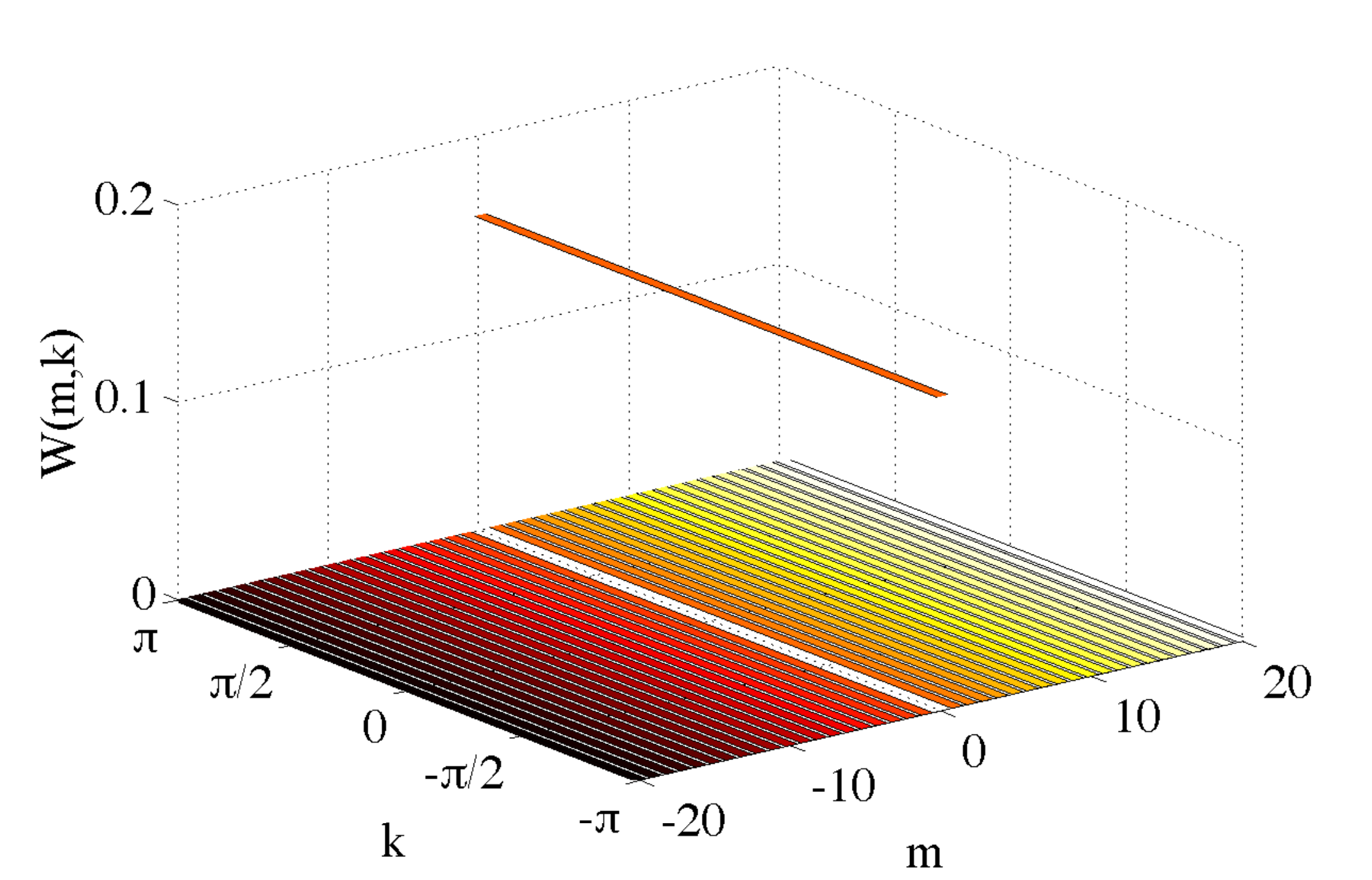}
\caption{Wigner function for a state localized at the origin $n_0=0$, assuming $a=1$. Notice that a plane projection of the phase space is represented, although it is periodic in $k$, and the edges $k=\pm \pi$ have to be identified.}
\label{fig:delta}
\end{figure}

This function, represented in \fref{fig:delta}, is non-negative everywhere, so that $\eta(\delta)=0$.

\subsection{Gaussian state}

In the case of continuous degrees of freedom, Gaussian states play a fundamental role. In particular, pure
Gaussian states are the only pure states with non negative Wigner function~\cite{hudson74}.
It makes then sense to consider the discretization of a state
$\Psi(x)=\frac{1}{(\sigma \sqrt{ \pi})^{\frac{1}{2}}}e^{-\frac{(x-x_0)^2}{2 \sigma^2}}e^{i q_0 x}$,
namely
 $|\Psi_G\rangle =\frac{1}{N} \sum_n e^{-\frac{(n-n_0)^2}{2 \ssigma^2}} e^{i q_0 n a } |n\rangle$, for $n_0\in\mathbb{Z}$ being $\ssigma\equiv\sigma/a$ the width measured in units of the lattice spacing.
The correct normalization in the discrete case, ${N}^2=\sum_ne^{-\frac{(n-n_0)^2}{\ssigma^2}}\equiv\theta_3(0,e^{-\frac{1}{\ssigma^2}})$, is expressed in terms of the Jacobi theta function, defined as $\theta_3(z,q)\equiv \sum_n q^{n^2} e^{2 i z n}$ for complex arguments $q$, $z$,
with $|q|<1$~\cite{abramsteg}.


\begin{figure}
\includegraphics[width=.6\columnwidth]{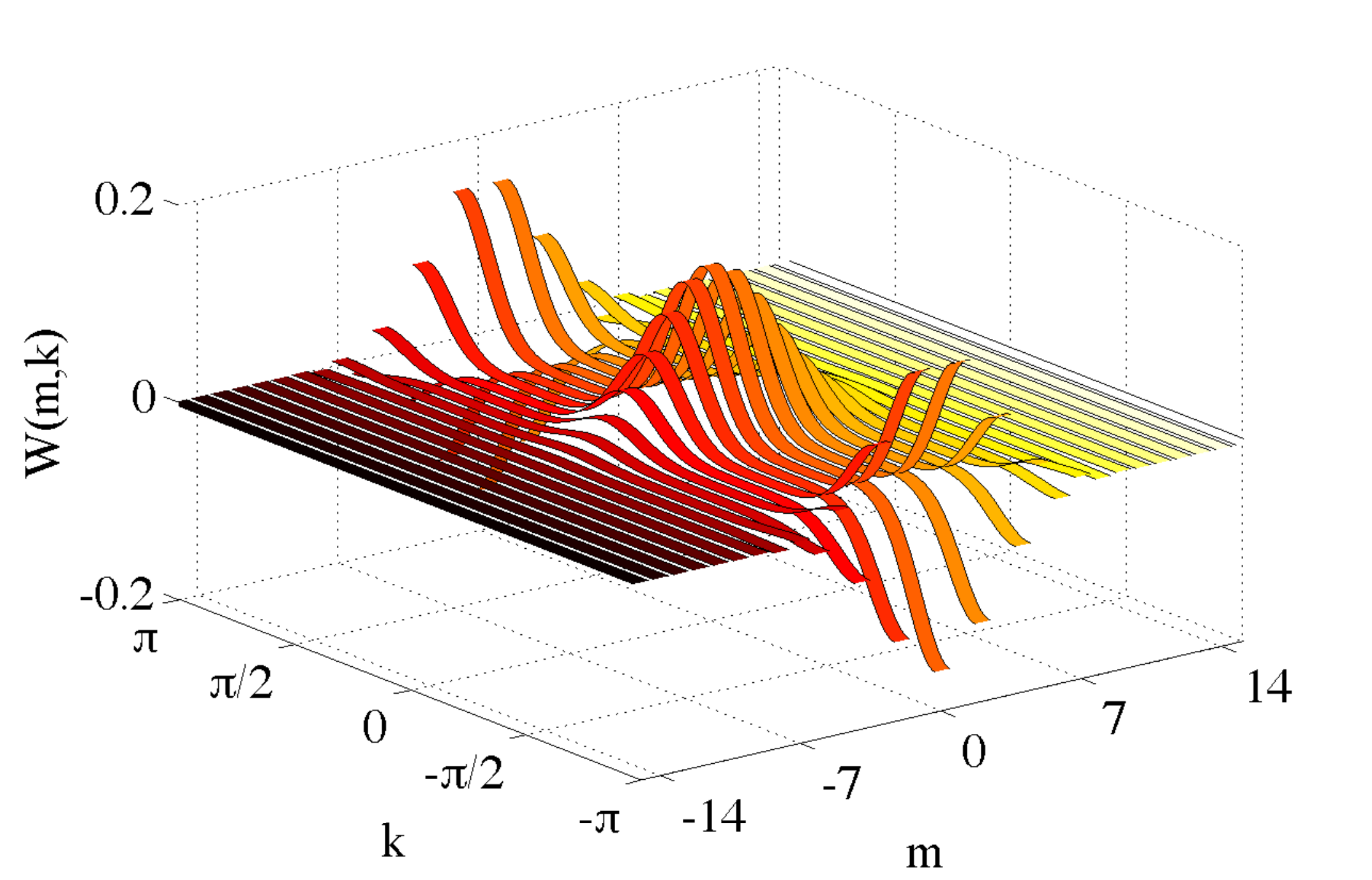}
\caption{Wigner function for a discretized Gaussian state with $\ssigma=2$, $q_0=0$ and $n_0=0$, taking $a=1$.}
\label{fig:gauss}
\end{figure}

\begin{figure}
\begin{minipage}[t]{.5\columnwidth}
\subfigure{
\label{fig:gaussDispl}
\includegraphics[width=\columnwidth]{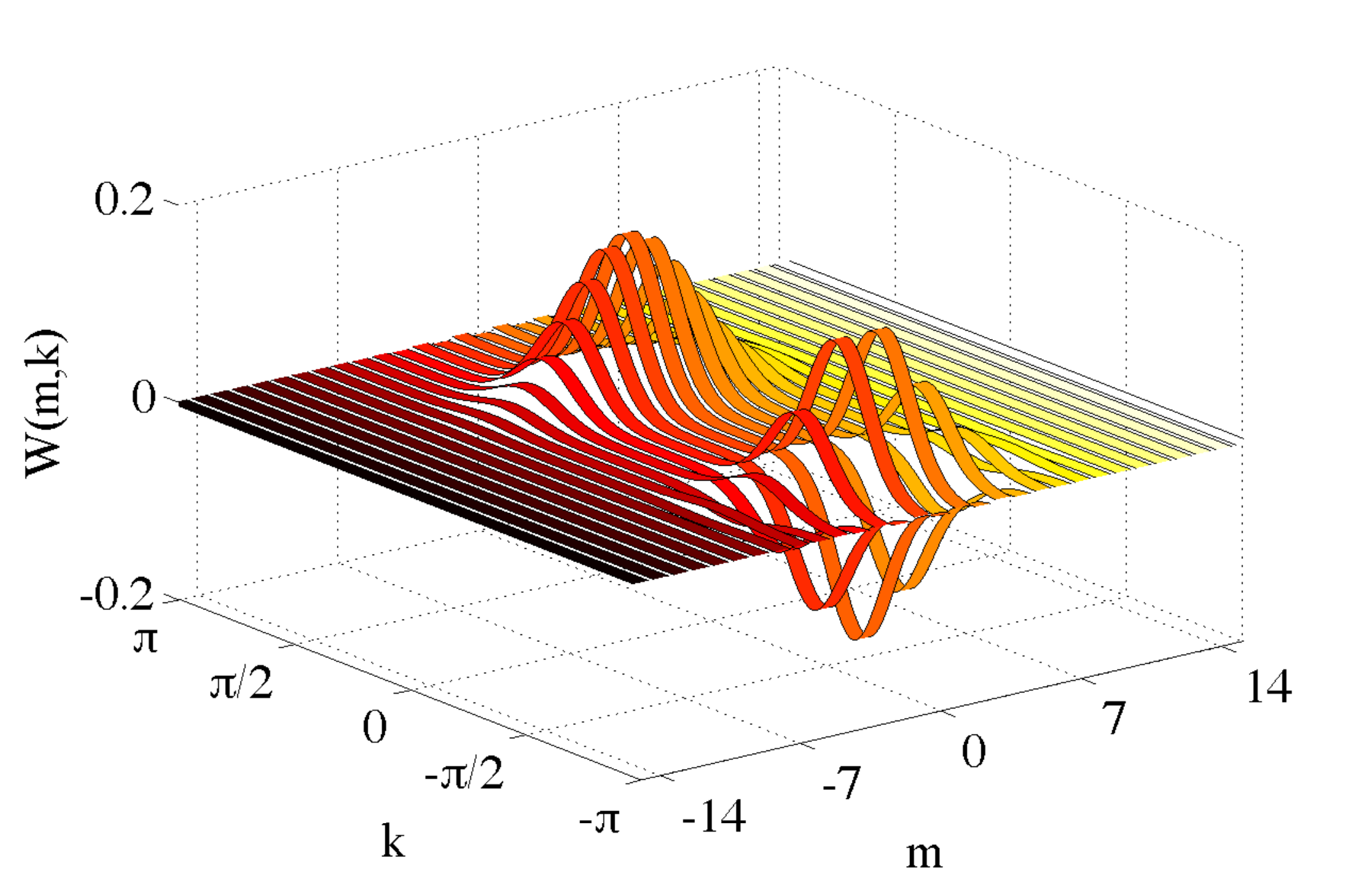}
}
\end{minipage}
\begin{minipage}[t]{.5\columnwidth}
\subfigure{
\label{fig:gaussWtilde}
\includegraphics[width=\columnwidth]{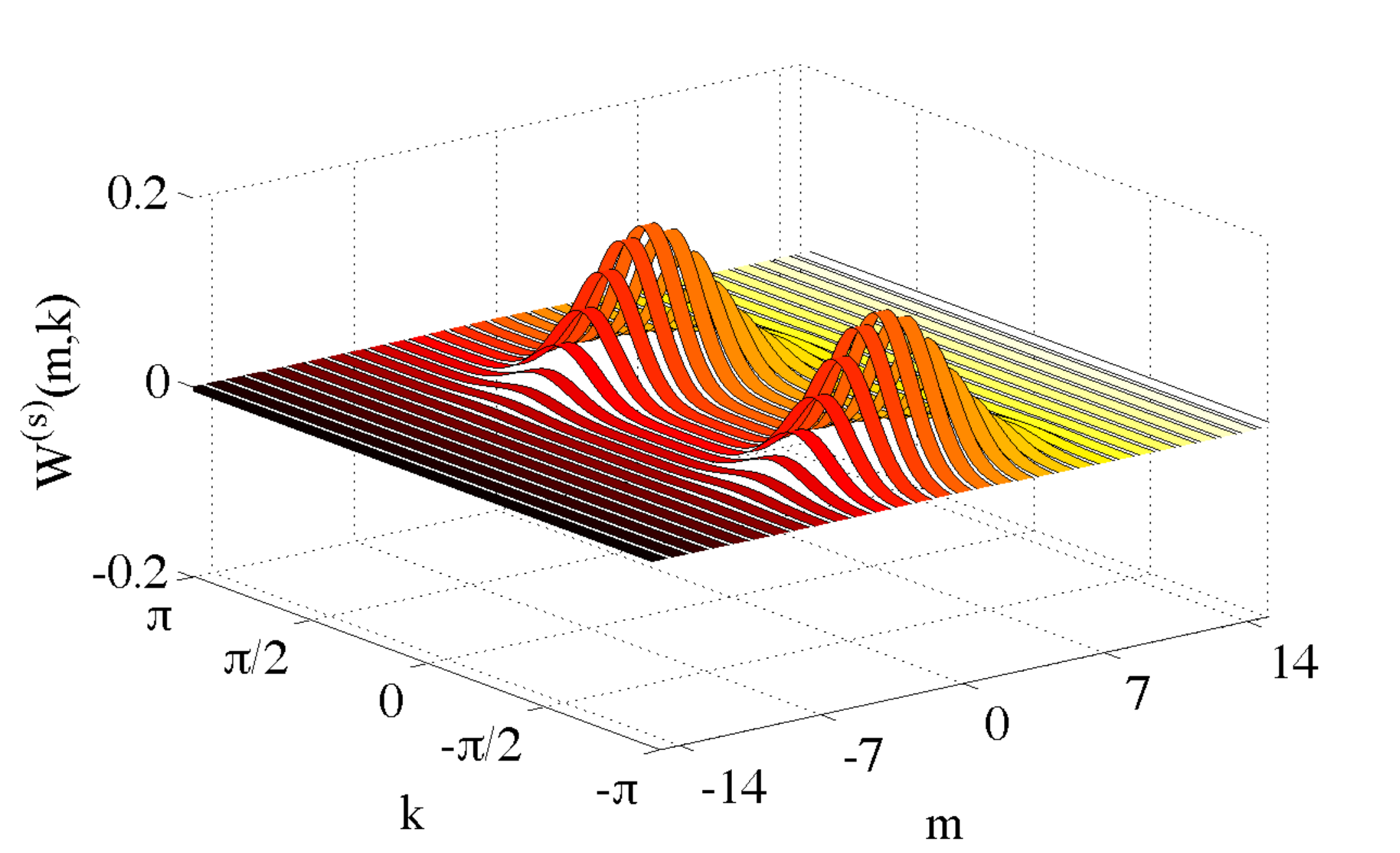}
}
\end{minipage}
\caption{Wigner function (left) for a discretized Gaussian state identical to that in \fref{fig:gauss} with a displacement in momentum $q_0 a=\pi/3$.
On the right, the sign-averaged Wigner function \eqref{eq:defWtilde}.}
\label{fig:gaussDisplaced}
\end{figure}


The Wigner function for this state can also be computed exactly, 
\beq
W_G(m,k)=\frac{1}{2\pi} e^{i(k-q_0 a)m} e^{-\frac{(m-n_0)^2+n_0^2}{2 \ssigma^2}}
\frac{\theta_3(k-q_0 a + i \frac{m}{2 \ssigma^2},e^{-\frac{1}{\ssigma^2}})}{ \theta_3(0,e^{-\frac{1}{\ssigma^2}})},
\label{eq:WFgauss}
\eeq
and shown in \fref{fig:gauss} for the particular case $\ssigma=2$, $n_0=0$, $q_0=0$.
The figure shows clearly the regular image, centered around $k=0$, and the \emph{ghost} image,
exhibiting the sign oscillations on odd sites.
If we consider instead a displaced Gaussian, with $q_0\neq 0$, the whole figure is correspondingly shifted in
momentum space, as shown in \fref{fig:gaussDispl}.
To illustrate the meaning of the sign-averaged function defined in \eqref{eq:defWtilde}, we also plot it in \fref{fig:gaussWtilde} for this state.
Obviously, $\eta(\Psi_G)=0$ for any pure  Gaussian state.


\subsection{Superposition of deltas}

The Gaussian case has vanishing negativity, as expected from its correspondence in the continuum limit. 
It actually includes the case of a localized state, too, which can be interpreted as a Gaussian in the limit of a vanishing width, $\ssigma$.
Superpositions of such states will instead have more \emph{quantum} features.

%

\begin{figure}
\includegraphics[width=.6\columnwidth]{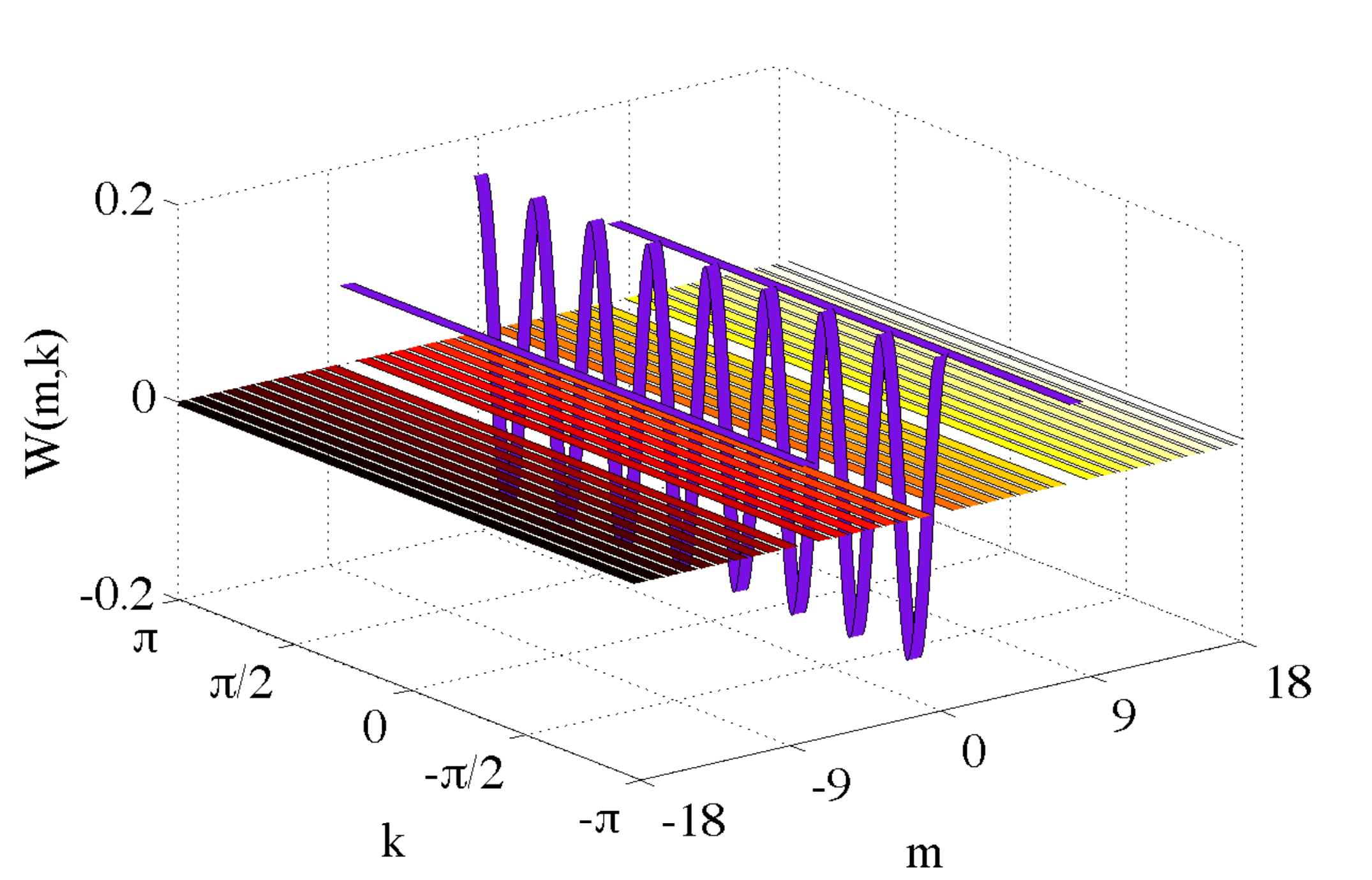}
\caption{Wigner function for the superposition of two deltas located at $n_1=-n_2=4$, assuming lattice spacing $a=1$. 
The function vanishes everywhere except on three isolated 
strips, colored purple in the figure.}
\label{fig:twodeltas} 
\end{figure}

We may in particular consider an arbitrary superposition of two localized states, such as
$| \Psi_{2\delta}\rangle=\frac{1}{\sqrt{1+|\alpha|^2}}\sum_n \left ( \delta_{n n_1} + \alpha \delta_{n n_2}\right )|n\rangle,$ for any $n_1\neq n_2 \in \mathbb{Z}$ and $\alpha\in\mathbb{C}$. The corresponding 
Wigner function can be easily calculated,
\bea
W_{2\delta}(m,k)&=& \frac{1}{2\pi (1+|\alpha|^2)} \left\{ \delta_{m,2n_1} + |\alpha|^2 \delta_{m,2 n_2} 
 \right. \nonumber \\ 
 && \quad \left.
 + 2|\alpha| \delta_{m, n_1+n_2}\cos[\Delta n \ k +\phi ] \right\} ,
\eea
where $\phi$ is the phase of the complex coefficient $\alpha$, and $\Delta n=n_2-n_1$.
In this case, the Wigner function vanishes everywhere except for three particular values of the space-like phase space coordinate, 
namely $m=2n_1,\, 2 n_2, \, n_1+n_2$. 
Figure \ref{fig:twodeltas}  shows the particular case of $n_1=-n_2=4$, $\alpha=1$.

It is easy to see that $W^{(s)}_{2\delta}(m,k)=W_{2\delta}(m,k)$, since none of the terms changes sign under 
\eqref{eq:defWtilde}.
Indeed, the first two terms are non-vanishing only on even values of $m$, while the last term
can be supported on odd $m$ if $n_1$ and $n_2$ have different parity, but in that case, $W_{2\delta}(m\pm 1,k)=0$.
Therefore, we can analytically compute the quantity \eqref{eq:neg} as
\bea
\eta(\Psi_{2\delta})&=& \sum_m \int_{-\pi}^{\pi}[|W(m,k)|-W(m,k)] dk
\nonumber \\
&=&\frac{ |\alpha|}{\pi (1+|\alpha|^2)} \int_{-\pi}^{\pi} dk \left[ |\cos(\Delta n k +\phi)|
-\cos(\Delta n k +\phi)\right]
\nonumber \\
&=&\frac{4 |\alpha|}{\pi (1+|\alpha|^2)},
\label{eq:neg2deltaB}
\eea
independent of the separation between the localized states, $\Delta n$, and 
reaching its maximum value, $\eta_{\mathrm{max}(\Psi_{2\delta})}=2/\pi$, for $|\alpha|=1$.


\subsection{Superposition of Gaussian states}

Another family of states for which the Wigner function defined above can be computed analytically is that of superpositions
of pure Gaussian states.
We may consider an arbitrary superposition of two discretized pure Gaussian states,
\beq
|\Psi_{2G}\rangle=\frac{1}{\cal N} \sum_n \left\{ e^{-\frac{(n-n_1)^2}{2 \ssigma_1^2}}e^{i q_1 n a}+ \alpha e^{-\frac{(n-n_2)^2}{2 \ssigma_2^2}}e^{i q_2 n a} \right\}|n\rangle,
\eeq
for arbitrary $n_{1,2}\in\mathbb{Z}$, $q_{1,2}\in[-\pi/a,\pi/a[$ and $\alpha\in\mathbb{C}$. For such state, the 
Wigner function can be expressed as a sum
\beq
W_{2G}=W_1+|\alpha|^2 W_2 + \alpha W_{12} + \alpha^* W_{21},
\eeq
where $W_1$ and $W_2$ are (up to the normalization factor) equivalent to the Wigner function of a single Gaussian \eqref{eq:WFgauss},
while $W_{12}$ and $W_{21}$ contain the crossed terms, 
\bea
W_{12}(m,k)&=&\frac{1}{\pi {\cal N}^2} e^{i(k-q_2a)m} e^{-\frac{n_1^2}{2 \ssigma_1^2}}e^{-\frac{(m-n_2)^2}{2 \ssigma_2^2}}
\nonumber \\
&&\times \theta_3\left(k-a \tfrac{q_1+q_2}{2}+i\left(\tfrac{m-n_2}{2\ssigma_2^2}+\tfrac{n_1}{2 \ssigma_1^2}\right),
e^{-\frac{\ssigma_1^2+\ssigma_2^2}{2 \ssigma_1^2 \ssigma_2^2}}\right),
\eea
and $W_{21}=W_{12}(1\leftrightarrow 2)$.

In the symmetric case, $\alpha=1$, $n_1=-n_2\equiv n_0$, $\ssigma_1=\ssigma_2\equiv\ssigma$, $q_1=q_2=0$,
the above expression adopts the compact form
\bea
W_{2G}(m,k)&=&
\frac{ e^{i k m }}{\pi {\cal N}^2}  e^{-\frac{m^2}{2 \ssigma^2}} \left \{
e^{-\frac{n_0^2}{\ssigma^2}}\mathrm{cosh}\frac{m n_0}{\ssigma^2}+\cos(2 k n_0)
\right \}
\nonumber \\
&&\times \theta_3 \left( k  +i\frac{m}{2 \ssigma^2},e^{-\frac{1}{\ssigma^2}}
\right),
\eea
with ${\cal N}^2=2(1+e^{-{n_0^2}/{\ssigma^2}})\theta_3(0,e^{-{1}/{\ssigma^2}})$.
Using the properties of the $\theta_3$ function, we can further simplify the expression, so that, for even $m=2 s$,
\beq
W_{2G}(2 s,k)=\frac{e^{-\frac{s^2}{\ssigma^2}} }{\pi {\cal N}^2 } \theta_3(k ,e^{-\frac{1}{\ssigma^2}})
\left \{
 e^{-\frac{n_0^2}{\ssigma^2}} \mathrm{cosh}\frac{2 s n_0}{\ssigma^2}+\cos(2 k n_0 )
\right\},
\label{eq:2Ge}
\eeq
and for odd $m=2s+1$,
\bea
W_{2G}(2 s+1,k)&=&\frac{e^{ik} }{\pi {\cal N}^2 } e^{-\frac{(s+1/2)^2}{\ssigma^2}} e^{-\frac{1}{4 \ssigma^2}}\theta_3(k+\frac{i}{2 \ssigma^2},e^{-\frac{1}{\ssigma^2}})
\nonumber \\
&&\times \left \{
 e^{-\frac{n_0^2}{\ssigma^2}} \mathrm{cosh}\frac{(2 s+1) n_0}{\ssigma^2}+\cos(2 k n_0)
\right\}.
\label{eq:2Go}
\eea

In the limit $\ssigma\to 0$, \eqref{eq:2Ge} results in the expression for the superposition of two localized states discussed in the 
previous section, while \eqref{eq:2Go} vanishes.

\begin{figure}
\begin{minipage}[t]{.5\columnwidth}
\subfigure{
\label{fig:twogauss}
\includegraphics[width=\columnwidth]{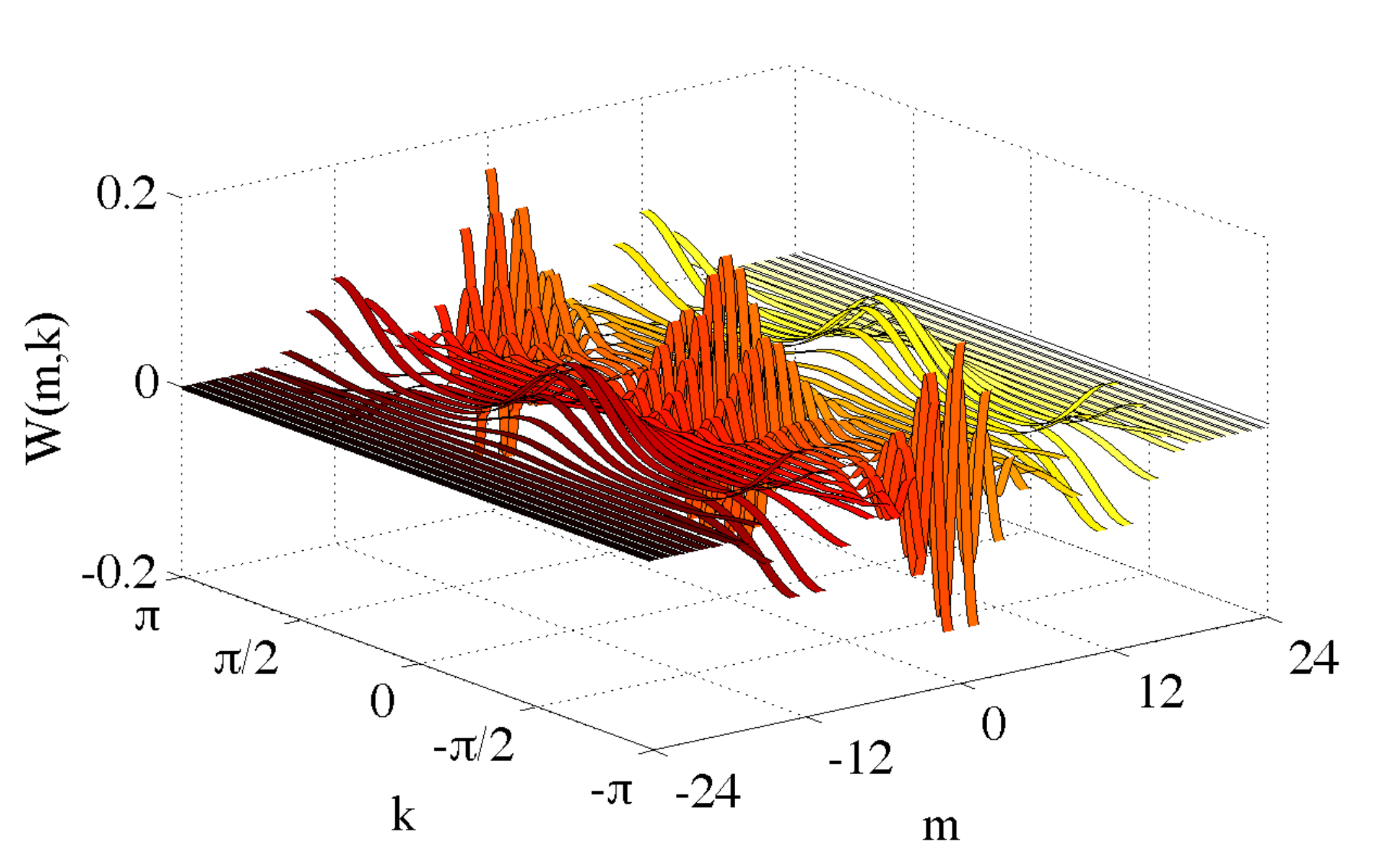}
}
\end{minipage}
\begin{minipage}[t]{.5\columnwidth}
\subfigure{
\label{fig:twogaussWtilde}
\includegraphics[width=\columnwidth]{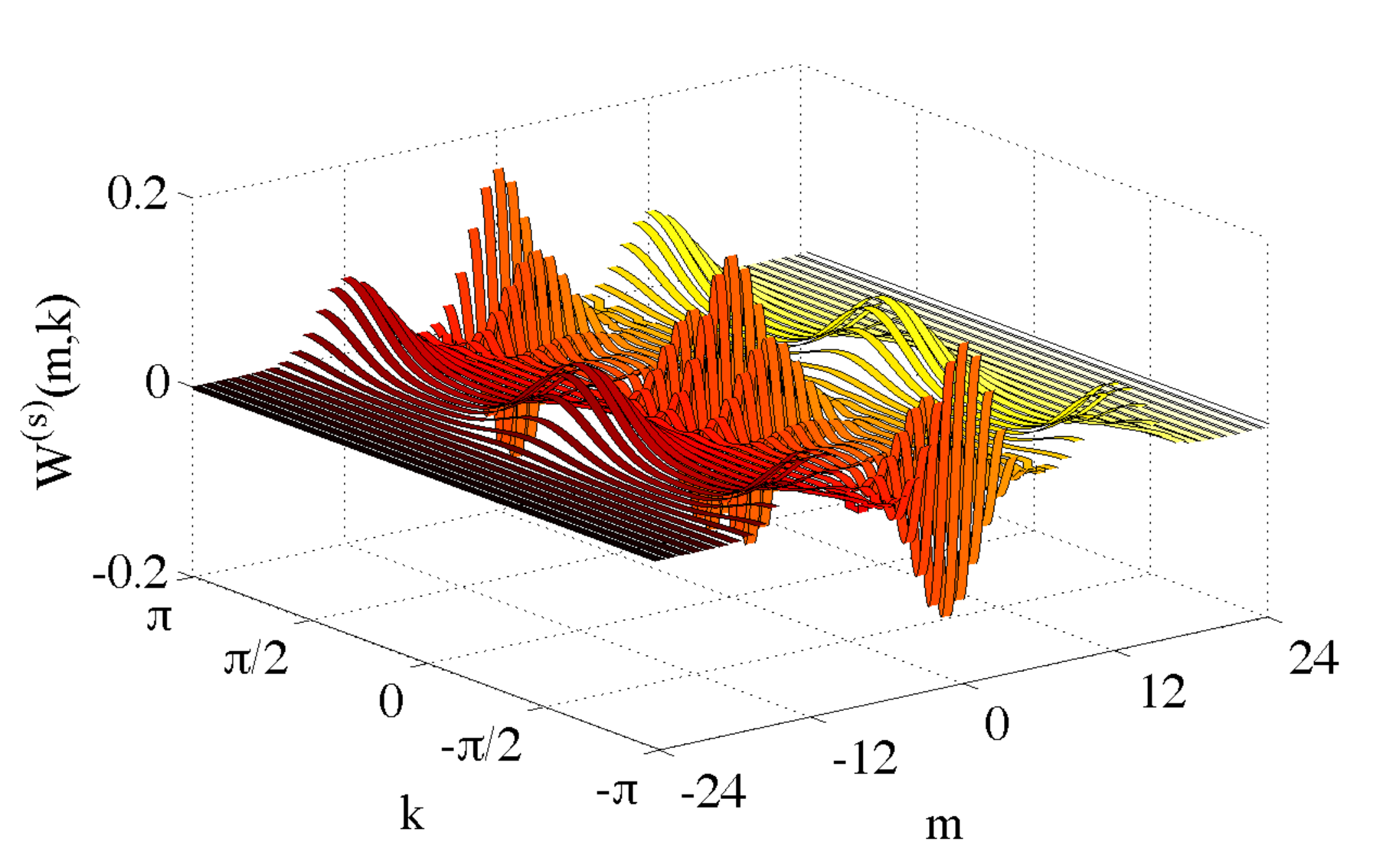}
}
\end{minipage}
\caption{Wigner function (left) for the symmetric superposition of two Gaussian states with center in $\pm n_0$, for $n_0=6$, and width $\ssigma=1.5$, assuming lattice spacing $a=1$. The right panel shows the sign-averaged $W^{(s)}$, for comparison.}
\label{fig:twogaussBoth}
\end{figure}

Figure \ref{fig:twogauss} shows the full Wigner function for the particular case $n_0=6$, $\ssigma=1.5$. The  
central part, around $k=0$, corresponds to the regular image, showing the usual Gaussian peaks and a central interference region.
This survives in the continuum limit, giving rise to a genuine negativity.
The ghost image in this case lives on the half phase-space with larger momenta, and exhibits the characteristic sign oscillation when
moving along the space-like axis. The sign average defined in \eqref{eq:defWtilde} transforms this image in a copy of the genuine one, as shown in \fref{fig:twogaussWtilde}, so that $\eta(\Psi_{2G})$ will be twice the negativity of the regular image.


\begin{figure}
\begin{minipage}[t]{.5\columnwidth}
\subfigure{
\label{fig:twogaussNega}
\includegraphics[width=\columnwidth]{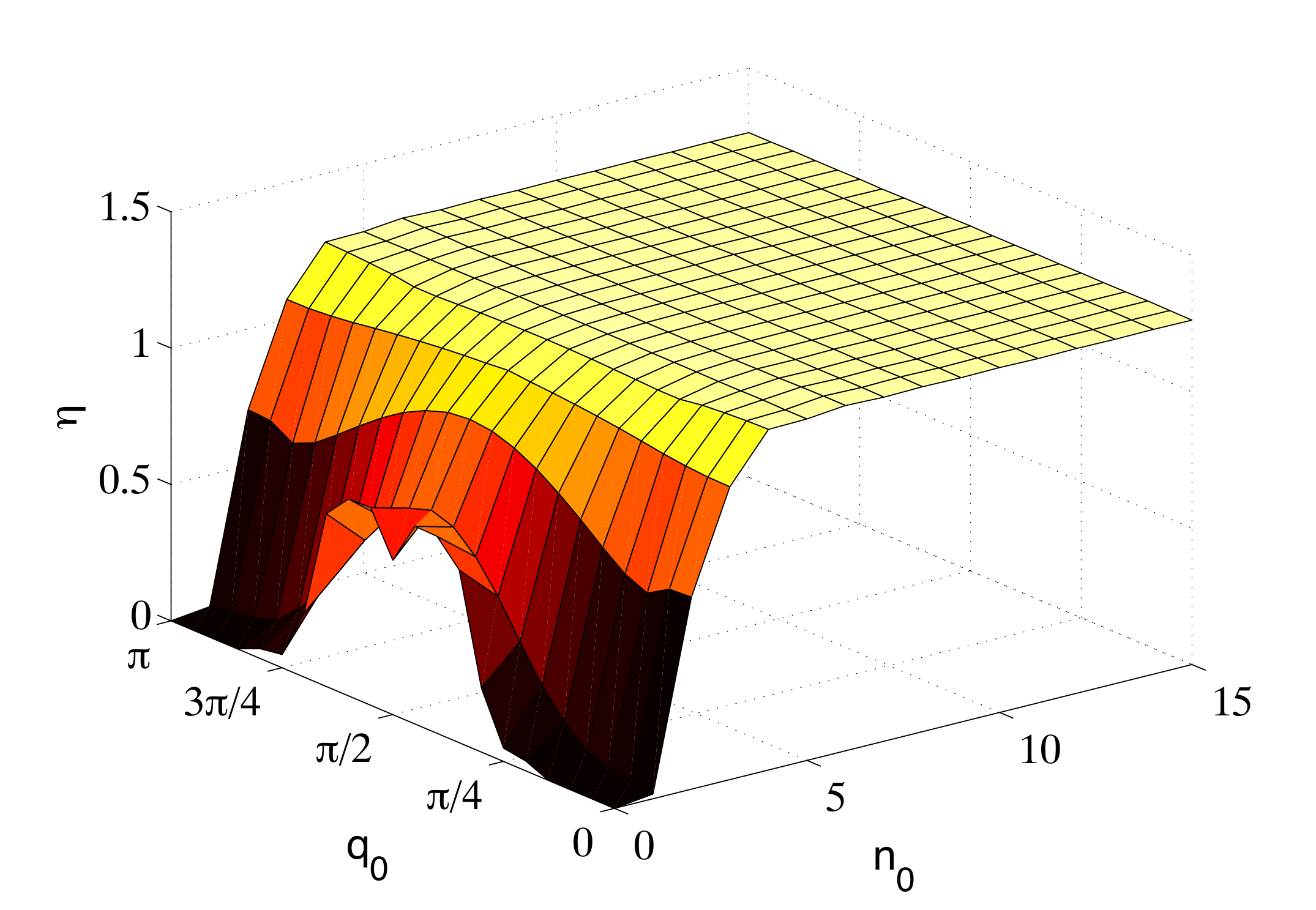}
}
\end{minipage}
\begin{minipage}[t]{.5\columnwidth}
\subfigure{
\label{fig:twogaussNegb}
\includegraphics[width=\columnwidth]{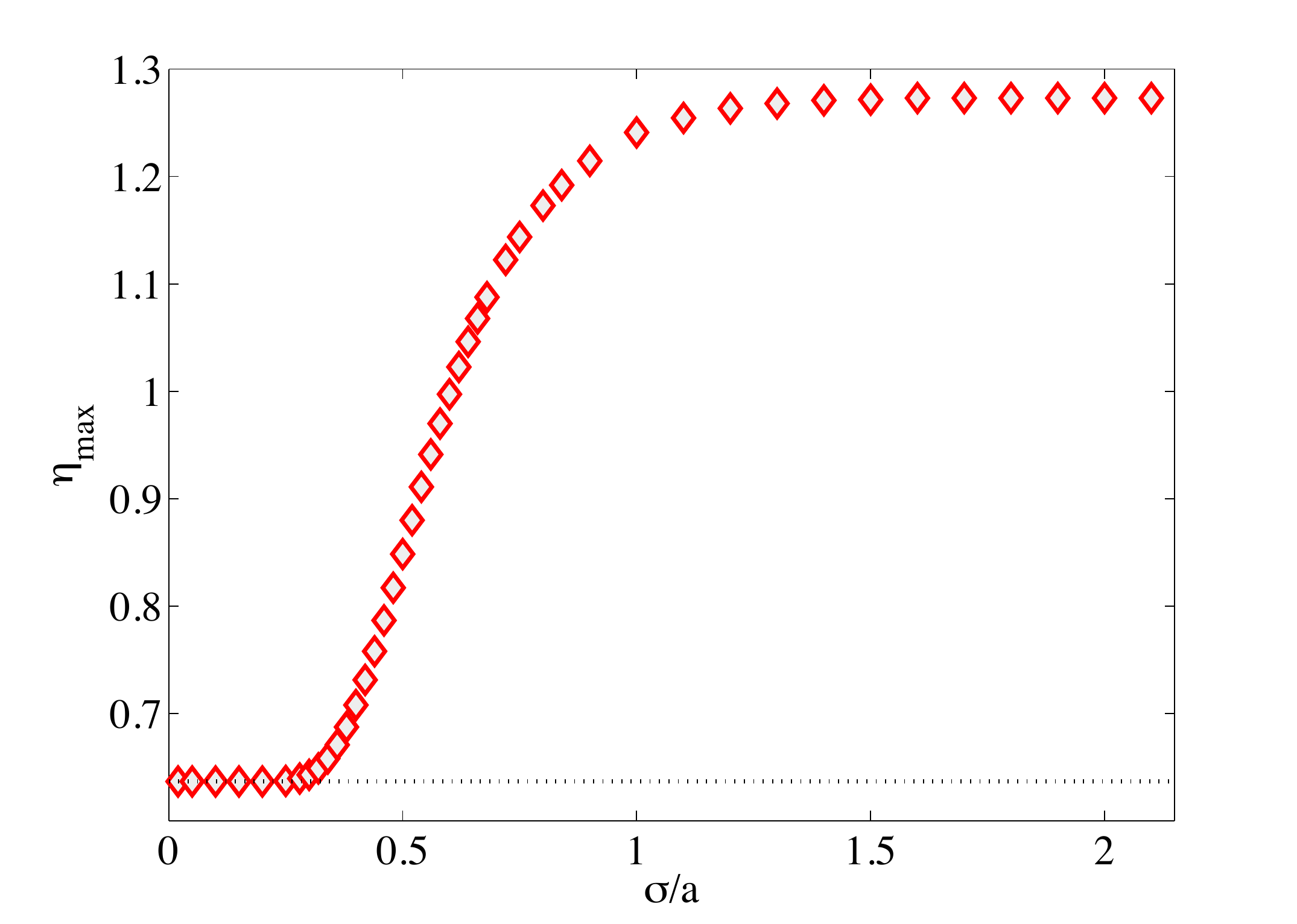}
}
\end{minipage}
\caption{Negativity of a superposition of two discretized Gaussian states of the same width. The left plot shows the case $\ssigma=1.2$, as a function of their half-distance, $n_0$, and the relative momentum displacement, $q_0$, taking $a=1$. On the right, we show the asymptotic value reached for varying width, $\ssigma$.}
\label{fig:twogaussNeg}
\end{figure}

Although there is no closed analytical expression for the non-classicality $\eta(\Psi_{2G})$, even in the simplest case discussed above, 
we can compute it numerically, as shown in \fref{fig:twogaussNega} for the symmetric superposition of two Gaussian states of the same width, 
centered at $\pm n_0$ and with momentum displacements, $q_1=0$ and $q_2\equiv q_0$. 
As shown in the plot, $\eta$ vanishes only for $n_0=0$ and $q_0=0,\,\pi$, when the situation reduces to a single Gaussian.
For small distances, $2 n_0$, the value of $\eta$ depends on 
 $n_0$ and $q_0$, while for larger separations it becomes less sensitive to $q_0$, and soon enough it reaches its maximal value, and 
 stays constant.
As shown in \fref{fig:twogaussNegb}, this asymptotic value is sensitive to the Gaussian width only when the latter is comparable to the lattice spacing.
When $\ssigma$ is large enough, instead, the asymptotic negativity is constant.
In the limit $\ssigma\to 0$, on the other hand, the negativity for a superposition of two deltas is recovered.

\section{\label{sec:discussion}Discussion}

We have extended the formalism of the Wigner function to the case of a quantum system with a discrete, infinite dimensional 
Hilbert space. For instance, this would be the case for a spinless particle moving on a one dimensional lattice.
The prescription presented here appears to be the natural one for this problem, as it satisfies the defining mathematical properties of the phase-space representation and recovers the correct continuum limit for vanishing lattice spacing.

The quantification of non-classicality, as signaled by the negative part of the Wigner function in the case of continuous degrees of freedom, has to be redefined in this case to exclude negative contributions due to the structure of the discrete phase space itself. We have proposed a negativity measure for this case, and have illustrated it with the explicit results for localized and Gaussian states, and for superpositions of each.
Our results support the meaningfulness of this measure to characterize the states of a particle on a one-dimensional lattice.

As for other cases, in which the phase-space formalism can also account for the dynamics of the system, it would be possible to formulate
the evolution of such system fully in terms of its Wigner function, and to
use the proposed measure, $\eta(\rho)$, to
classify quantumness in evolving states.
Although the examples presented in this paper are focused on pure states, the same concepts apply also to 
mixed states.

An interesting extension of this work is combining the phase space introduced here with additional degrees of freedom,
such as internal ones for the particle, or to extend it to the case of several particles or dimensions.
Wootter's prescription \cite{Wootters1987} to construct composite phase spaces by combining 
the phase-space point operators of different degrees of freedom via their tensor product can be applied in this case.

\begin{acknowledgments}
This work was partly funded by the Spanish Grants FPA2011-23897 and \emph{Generalitat
Valenciana} grant PROMETEO/2009/128, and by the DFG by Forschungsgruppe 635.

We gratefully acknowledge the support of \emph{Centro de Ciencias Pedro Pascual} in Benasque (Spain),
where part of this work was developed.
\end{acknowledgments}

\appendix
\section{Pure states with positive Wigner function}

Analogous to the result in~\cite{rigas10} for a conjugate pair of angle and angular momentum variables, with the present definition 
the Wigner function of a pure state is non negative if and only if it is an eigenstate of the discrete position operator, i.e.
$\langle n|\Psi\rangle=\delta_{n\, n_0}$.
The first part of the theorem is trivial, since the Wigner function of a localized state \eqref{eq:WFdelta} 
is non negative.

To show the converse, let us assume a pure state with non-negative Wigner function, $W(m,k)\geq0,$
$\forall m\in\mathbb{Z}, \, k\in[-\pi,\pi[$.
From \eqref{eq:phases} it follows that the Wigner function can only be non-vanishing on points of the phase space with even space-like
coordinate, $m=2 n$,
\begin{equation}
W(2n+1,k)=0 \quad \forall n\in\mathbb{Z}.
\label{eq:app1}
\end{equation}

The rest of the demonstration follows closely that in~\cite{rigas10}, 
and we sketch it here only for completeness, 
with the proper modifications to match the
definition in \eqref{eq:WFdef}. 

The proof relies on the following two lemmas, proven in  \cite{rigas10}, for complex periodic functions 
and their (discrete) Fourier transform.
\begin{enumerate}
\item{\label{lemma1}}
Let $g(q)$ be a continuous, complex, $2\pi-$periodic function.
If its Fourier transform is non-negative, then the integration kernel $g(q-q')$ is
non-negative.
\item{\label{lemma2}}
Given a function $f:\mathbb{Z}\rightarrow\mathbb{C}$, if its inverse
Fourier transform has constant modulus, then 
$\sum_{n\mathbb{\in Z}}f(n) f^{*}(n+m)=0$, $\forall m\neq0$.
\end{enumerate}

It is easy to see that, for a pure state, the Wigner function can be
written as
\beq
W(m,k)=\frac{1}{2 \pi}\int_{-\pi/a}^{\pi/a}dq e^{iqma}\tilde{\psi}(\tfrac{k}{a}+q)\tilde{\psi}^{*}(\tfrac{k}{a}-q)
\eeq
 where $\tilde{\psi}(k)=\langle k|\psi\rangle$ are the components of the state in the quasi-momentum basis.
It is thus the Fourier transform of the function $g(q)=\frac{1}{a\sqrt{2\pi}}\tilde{\psi}((k+q)/a)\tilde{\psi}^{*}((k-q)/a)$.
From lemma~\ref{lemma1}, $\int_{-\pi}^{\pi} dq' \chi^*(q) g(q-q') \chi(q') \geq 0$ for any $\chi$.
In particular, requiring the inequality for all functions ${\chi}(q)=a_{1}\delta_{2\pi}(q-c_{1})+a_{2}\delta_{2\pi}(q-c_{2})$,
where $\delta_{2\pi}(q)\equiv\sum_{r\in\mathbb{Z}}\delta(q-2 r\pi)$, $a_{1,2}\in\mathbb{C}$, $c_{1,2}\in\mathbb{R}$,
implies
\beq
|\tilde{\psi}(q)|^{2}\geq|\tilde{\psi}(q+\Delta)||\tilde{\psi}(q-\Delta)|, \quad \forall \Delta\in\mathbb{R}, \quad q\in[-\tfrac{\pi}{a},\tfrac{\pi}{a}[.
\eeq
This requires that 
$|\tilde{\psi}(q)|$ is constant, so that also $|g(q)|$ must
be constant. Applying now lemma~(\ref{lemma2})
\beq
\sum_{j}W(m,k)W(m+j,k)=0 \quad \forall  j\neq0.
\eeq
So that, for a given value of $k$, there can at most be a single space-like component, $m_0(k)$, 
for which the Wigner function does not vanish.
Combining this with \eqref{eq:app1}, we obtain that such component must be even, $m_0(k)=2 n_0(k)$,
so that, using normalization,
 $W(m,k)=\frac{1}{2\pi}\delta_{m,2 n_{0}(k)}$.

It only remains to be shown that this component is the same for all values of $k$.
This can be seen, as in~\cite{rigas10}, by making use of the expression for the
Wigner function for a product in terms of individual Wigner functions,
\bea
W_{\varrho_{2}\varrho_{1}}(m,k)=\frac{1}{2\pi}\sum_{m_{1},m_{2}}\int_{-\pi}^{\pi}
 dk_{1}\int_{-\pi}^{\pi} dk_{2} &&W_{\varrho_{1}}(m+m_{1},k+k_{1})
\\
&&
W_{\varrho_{2}}(m+m_{2},k+k_{2})  e^{i (m_{2}k_{1}-m_{1}k_{2})}.
\nonumber 
\eea
In particular, taking $\rho_1=\rho_2\equiv \rho,$ the pure state we are considering, for which 
$W(m,k)=\frac{1}{2\pi}\delta_{m,2 n_{0}(k)}$, and looking at the (real) component for $m=2 n_0(0)\equiv 2 n_0$, $k=0$,
\beq
4 \pi^2=\int_{-\pi}^{\pi} d k_1 \int_{-\pi}^{\pi} d k_2
\cos \left ( 2 k_2 [n_0(k_1)-n_0]-2 k_1[n_0(k_2)-n_0]
\right ).
\eeq
To fulfill this equality, the argument of the cosine has to be an integer multiple of $2\pi$ for all values of $k_{1,2}$,
which is only possible if
$n_0(k)=n_0$ $\forall k$, and thus
\beq
W(m,k)=\frac{1}{2\pi}\delta_{m,2n_{0}}.
\eeq
Using \eqref{eq:rhoWF} it is easy to show that the pure state corresponding to this Wigner function is $|\Psi_{\delta}\rangle=|n_0\rangle$.
\qed

\bibliography{qwalks,wigner}

\begin{thebibliography}{10}

\bibitem{Hillery1984121}
M.~Hillery, R.~F. O'Connell, M.~O. Scully, and E.~P. Wigner.
\newblock Distribution functions in physics: Fundamentals.
\newblock {\em Physics Reports}, 106(3):121--167, 1984.

\bibitem{ferrie11review}
Christopher Ferrie.
\newblock Quasi-probability representations of quantum theory with applications
  to quantum information science.
\newblock {\em Reports on Progress in Physics}, 74(11):116001, 2011.

\bibitem{PhysRev.40.749}
E.~Wigner.
\newblock On the quantum correction for thermodynamic equilibrium.
\newblock {\em Phys. Rev.}, 40(5):749--759, Jun 1932.

\bibitem{moyal49}
Jose Moyal.
\newblock Quantum mechanics as a statistical theory.
\newblock {\em Proc. Camb. Phil. Soc.}, 45:99--124, 1949.

\bibitem{bertrand87}
J.~Bertrand and P.~Bertrand.
\newblock A tomographic approach to wigner's function.
\newblock {\em Foundations of Physics}, 17:397--405, 1987.

\bibitem{vogel89}
K.~Vogel and H.~Risken.
\newblock Determination of quasiprobability distributions in terms of
  probability distributions for the rotated quadrature phase.
\newblock {\em Phys. Rev. A}, 40:2847--2849, Sep 1989.

\bibitem{citeulike:4313181}
Wolfgang~P. Schleich.
\newblock {\em {Quantum Optics in Phase Space}}.
\newblock Wiley-VCH, 1 edition, February 2001.

\bibitem{kenfack}
Anatole Kenfack and Karol Zyczkowski.
\newblock Negativity of the wigner function as an indicator of
  non-classicality.
\newblock {\em Journal of Optics B: Quantum and Semiclassical Optics},
  6(10):396, 2004.

\bibitem{stratonovich}
R.~L. Stratonovich.
\newblock On distributions in representation space.
\newblock {\em J. Exp. Theor. Phys. 4}, pages 891--898, 1957.

\bibitem{Wootters1987}
William~K. Wootters.
\newblock A wigner-function formulation of finite-state quantum mechanics.
\newblock {\em Annals of Physics}, 176(1):1 -- 21, 1987.

\bibitem{PhysRevA.70.062101}
Kathleen~S. Gibbons, Matthew~J. Hoffman, and William~K. Wootters.
\newblock Discrete phase space based on finite fields.
\newblock {\em Phys. Rev. A}, 70(6):062101, Dec 2004.

\bibitem{leonhardt95prl}
Ulf Leonhardt.
\newblock Quantum-state tomography and discrete wigner function.
\newblock {\em Phys. Rev. Lett.}, 74:4101--4105, May 1995.

\bibitem{PhysRevA.53.2998}
Ulf Leonhardt.
\newblock Discrete wigner function and quantum-state tomography.
\newblock {\em Phys. Rev. A}, 53(5):2998--3013, May 1996.

\bibitem{miquel02qc}
C\'esar Miquel, Juan~Pablo Paz, and Marcos Saraceno.
\newblock Quantum computers in phase space.
\newblock {\em Phys. Rev. A}, 65:062309, Jun 2002.

\bibitem{galvao05speedup}
Ernesto~F. Galv\~ao.
\newblock Discrete wigner functions and quantum computational speedup.
\newblock {\em Phys. Rev. A}, 71:042302, Apr 2005.

\bibitem{cormick06class}
Cecilia Cormick, Ernesto~F. Galv\~ao, Daniel Gottesman, Juan~Pablo Paz, and
  Arthur~O. Pittenger.
\newblock Classicality in discrete wigner functions.
\newblock {\em Phys. Rev. A}, 73:012301, Jan 2006.

\bibitem{gross07prime}
D.~Gross.
\newblock Non-negative wigner functions in prime dimensions.
\newblock {\em Applied Physics B: Lasers and Optics}, 86:367--370, 2007.

\bibitem{vourdas04review}
A.~Vourdas.
\newblock Quantum systems with finite hilbert space.
\newblock {\em Reports on Progress in Physics}, 67(3):267, 2004.

\bibitem{wootters86mub}
W.~K. Wootters.
\newblock Quantum mechanics without probability amplitudes.
\newblock {\em Foundations of Physics}, 16(4):391--405, 1986.

\bibitem{PhysRevA.72.012309}
Juan~Pablo Paz, Augusto~Jos\'e Roncaglia, and Marcos Saraceno.
\newblock Qubits in phase space: Wigner-function approach to quantum-error
  correction and the mean-king problem.
\newblock {\em Phys. Rev. A}, 72(1):012309, Jul 2005.

\bibitem{1751-8121-44-34-345305}
J~Zak.
\newblock Doubling feature of the wigner function: finite phase space.
\newblock {\em Journal of Physics A: Mathematical and Theoretical},
  44(34):345305, 2011.

\bibitem{PhysRevA.68.052305}
Cecilia~C. L\'opez and Juan~Pablo Paz.
\newblock Phase-space approach to the study of decoherence in quantum walks.
\newblock {\em Phys. Rev. A}, 68:052305, Nov 2003.

\bibitem{paz02mixed}
Juan~Pablo Paz.
\newblock Discrete wigner functions and the phase-space representation of
  quantum teleportation.
\newblock {\em Phys. Rev. A}, 65:062311, Jun 2002.

\bibitem{hudson74}
R.L. Hudson.
\newblock When is the wigner quasi-probability density non-negative?
\newblock {\em Reports on Mathematical Physics}, 6(2):249 -- 252, 1974.

\bibitem{mandilara09}
A.~Mandilara, E.~Karpov, and N.~J. Cerf.
\newblock Extending hudson's theorem to mixed quantum states.
\newblock {\em Phys. Rev. A}, 79:062302, Jun 2009.

\bibitem{PhysRevLett.106.010403}
A.~Mari, K.~Kieling, B.~Melholt Nielsen, E.~S. Polzik, and J.~Eisert.
\newblock Directly estimating nonclassicality.
\newblock {\em Phys. Rev. Lett.}, 106:010403, Jan 2011.

\bibitem{gross06hudson}
David Gross.
\newblock Hudson's theorem for finite-dimensional quantum systems.
\newblock {\em J. Math. Phys.}, 47:122107, 2006.

\bibitem{rigas10}
I.~Rigas, L.~L. S\'anchez-Soto, A.~B. Klimov, J.~\ifmmode \check{R}\else
  \v{R}\fi{}eh\'a\ifmmode~\check{c}\else \v{c}\fi{}ek, and Z.~Hradil.
\newblock Non-negative wigner functions for orbital angular momentum states.
\newblock {\em Phys. Rev. A}, 81:012101, Jan 2010.

\bibitem{rigas11}
I.~Rigas, L.L. S{\'a}nchez-Soto, A.B. Klimov, J.~{\v{R}}eh{\'a}{\v{c}}ek, and
  Z.~Hradil.
\newblock Orbital angular momentum in phase space.
\newblock {\em Annals of Physics}, 326(2):426 -- 439, 2011.

\bibitem{peyrin86}
F.~Peyrin and R.~Prost.
\newblock A unified definition for the discrete-time, discrete-frequency, and
  discrete-time/frequency wigner distributions.
\newblock {\em Acoustics, Speech and Signal Processing, IEEE Transactions on},
  34(4):858 -- 867, aug 1986.

\bibitem{arguelles05}
Arturo Arg{\"u}elles and Thomas Dittrich.
\newblock Wigner function for discrete phase space: Exorcising ghost images.
\newblock {\em Physica A: Statistical Mechanics and its Applications},
  356(1):72--77, 2005.

\bibitem{abramsteg}
Milton {Abramowitz} and Irene~A. {Stegun}.
\newblock {\em Handbook of Mathematical Functions with Formulas, Graphs, and
  Mathematical Tables}.
\newblock Dover, New York, 1972.

\end{thebibliography}

\end{document}